\documentclass[twocolumn]{aa}
\usepackage{graphicx}
\usepackage{natbib}
\bibpunct{(}{)}{;}{a}{}{,}

\usepackage{txfonts}
\begin{document}
   \title{Reopening the TNOs Color Controversy: Centaurs Bimodality
and TNOs Unimodality}

   \titlerunning{Reopening the TNOs Color Controversy}

   \author{N.~Peixinho\inst{1,2}
          \and
          A.~Doressoundiram\inst{1}
          \and
          A.~Delsanti\inst{1}
          \and
          H.~Boehnhardt\inst{3}
          \and
          M.~A.~Barucci\inst{1}
          \and
          I.~Belskaya\inst{1,4}
          }

   \offprints{N. Peixinho}

   \institute{LESIA, Observatoire de Paris, 5 pl Jules Janssen, 
     F-92195 Meudon cedex, France\\ 
     \email{Nuno.Peixinho,Alain.Doressoundiram,Audrey.Delsanti,Antonella.Barucci@obspm.fr}
         \and
         CAAUL, Observat\'orio Astron\'omico de Lisboa, Tapada da Ajuda, 
         PT-1349-018 Lisboa, Portugal 
         \and
              Max-Planck-Institut f\"ur Astronomie, K\"onigstuhl 17,
               69117 Heidelberg, Germany\\
               \email{hboehnha@mpia-hd.mpg.de}
          \and
         Astronomical Observatory of Kharkov National University,
         Sumskaya St., 35, Kharkov, 61022, Ukraine\\ 
         \email{irina@astron.kharkov.ua}
              }

   \date{Received 6 August, 2003; accepted 12 September, 2003}

   \abstract{We revisit the Trans--Neptunian Objects (TNOs) color
controversy allegedly solved by \cite{TegRom03}. We debate the
statistical approach of the quoted work and discuss why it can not
draw the claimed conclusions, and reanalyze their data sample with a
more adequate statistical test. We find evidence for the existence of
two color groups among the Centaurs. Therefore, mixing both centaurs
and TNOs populations lead to the erroneous conclusion of a global
bimodality, while there is no evidence for two color groups in the
TNOs population alone. We use quasi--simultaneous visible color
measurements published for 20 centaurs (corresponding to about half of
the identified objects of this class), and conclude on the existence
of two groups. With the surface evolution model of \cite{Del03} we
discuss how the existence of two groups of Centaurs may be compatible
with a continuous TNOs color distribution.

   }

   \maketitle
%
%________________________________________________________________

\section{Introduction}

A large population of small icy bodies exists beyond the orbit of
Neptune.  First speculated by \cite{Leo30,Edg43,Edg49} and
\cite{Kui51}, their existence was observationally confirmed by
\cite{JewLuu93} only eleven years ago. Considered as remnants from the
formation of the solar system, these Trans--Neptunian Objects (TNOs)
constitute the Edgeworth--Kuiper Belt (EKB), therefore also frequently
known as Kuiper Belt Objects (KBOs). Currently, more than 700 of them
have been detected.
A different class of objects, the Centaurs, was found by
\cite{KowGeh77}; to date, more than 40 of these objects are known.
Orbiting mainly between Jupiter and Neptune, a strict dynamical
definition does not exist and frequently some objects are both classified 
as Centaurs and Scattered Disk Objects (SDOs). 
Centaurs are believed
to be ex--TNOs in a transition phase between the EKB and the
Jupiter--family comets \citep{Fer80,DunQuiTre88}. Their exact origin
location in the EKB has not yet been identified. While it is currently
believed Centaurs originate from the Scattered Disk Objects
\citep{DunLev97} --- a ``fuzzy'' subclass of TNOs with highly eccentric
and inclined orbits --- an origin in the Plutinos (TNOs in a 3:2 mean
motion orbital resonance with Neptune) has also been hypothesized
\citep{YuTre99}.

Since the very beginning of the first photometric measurements of
these objects their visible color distribution always been
very controversial. \cite{TegRom98,TegRom00} -- hereafter TR98 and TR00,
respectively -- reported the identification of two separated color
groups (blue--gray, {\it i.e.}, solar like, and very red) from samples
of 16 and 37 objects respectively. Other groups stated a continuous
color spreading such as the precursor works by
\cite{LuuJew96,Gre97,Bar99}, and subsequent studies \citep[see the
review by][for a summary]{Dor04}.
\cite{JewLuu01} analyzed the statistical significance of this
bimodality on TR98 and TR00 data samples and also on their own dataset
of 28 objects. They concluded that both results were formally
consistent with derivation from a uniform color distribution, {\it
i.e.} unimodality.
With 21 objects, measured in BVRIJ bands, \cite{Bar01} reported the
identification of four groups with a quasi--continuous
spreading. However this grouping is seen only with $V-I$ and $V-J$
color information and becomes marginal with $B-V$ and $V-R$ colors.
Finally, \cite{TegRom03} -- hereafter TR03 -- claimed the resolution
of the TNOs color controversy, now with an enlarged sample of 55
objects (50 objects with measured $B-V$ and $V-R$ colors, and 5 with 
$V-R$ only). 
Furthermore, with Monte Carlo simulations they showed that
with typical error bars of $0.05$ magnitude other teams would not see
any clear ``gap'' between their two detected groups, concluding that
observational methodologies leading to smaller error bars are the basis
for these color groups findings.

Whereas it is physically difficult to understand the existence of two
groups of colors, several explanations have been proposed to describe
the continuous color range of TNOs. First, TNOs might have real
intrinsic differences. \cite{Gom03} proposed a migration model
explaining the present composition of the EKB as the mixing of bodies
formed in very different parts of the Solar System. \cite{LuuJew96}
explored a surface evolution mechanism: a competition between
reddening space weathering and bluishing collisional resurfacing,
which causes the continuous color spreading. However this model
predicts a surface color variation with the objects rotation ({\em
e.g.} within a few hours) that has never been observed to date
\citep{JewLuu01}; it also implies color--orbital parameters
correlations very different from observations
\citep{TheDor03}. Nevertheless, modeling is still in progress:
\cite{Gil02}, with a more detailed space weathering process, proposed
a new version of this mechanism but of difficult testing. \cite{Del03}
revisited the model by \cite{LuuJew96} testing an additional
resurfacing process: cometary activity, which is compatible with a
surface homogeneity.

%__________________________________________________________________

\section{Two color groups' analysis}
%__________________________________________________________________

%- - - - - - - - - - - - - - - - - - - - - - - - - - - - - - - - - - -
   \begin{figure}
   \centering
   \includegraphics[width=8.2cm, bb= 50 50 342 302]{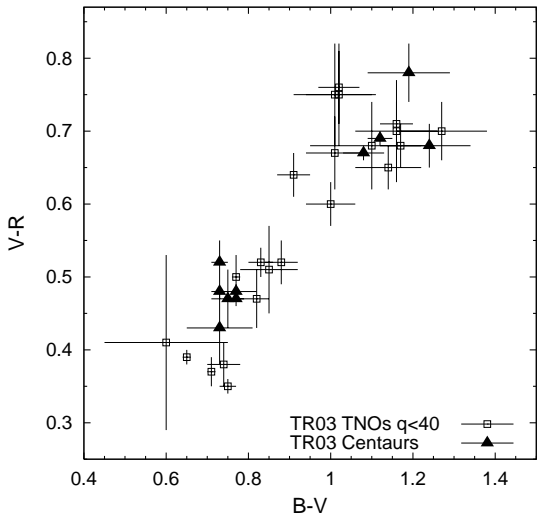}
   \caption[]{$B-V$ {\it vs.} $V-R$ plot of TR03's data used to conclude
for bimodality. Centaurs are highlighted and their bimodal behavior is
obvious, contrarily to TNOs taken alone.}
              \label{fig:tr03_bimod}
    \end{figure}
%- - - - - - - - - - - - - - - - - - - - - - - - - - - - - - - - - - -

%- - - - - - - - - - - - - - - - - - - - - - - - - - - - - - - - - - -
   \begin{figure}
   \centering
   \includegraphics[width=8.2cm, bb= 50 50 342 302]{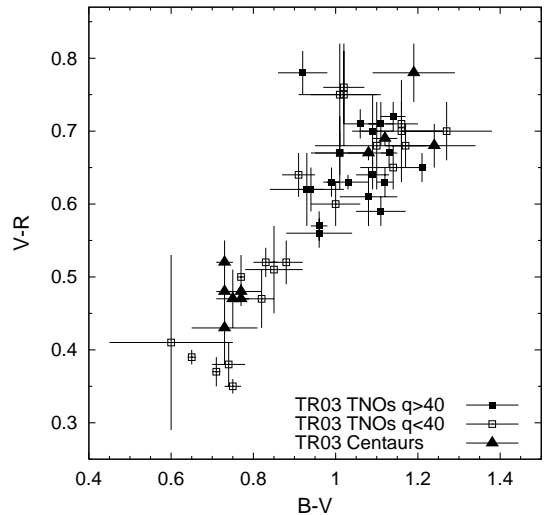}
      \caption[]{$B-V$ {\it vs.} $V-R$ plot of TR03's full data. The
``red cluster'' of TNOs with $q>40~AU$ is included.}
         \label{fig:tr03_full_data}
   \end{figure}
%- - - - - - - - - - - - - - - - - - - - - - - - - - - - - - - - - - -

%- - - - - - - - - - - - - - - - - - - - - - - - - - - - - - - - - - -
   \begin{figure}
   \centering
   \includegraphics[width=8.2cm, bb= 50 50 342 302]{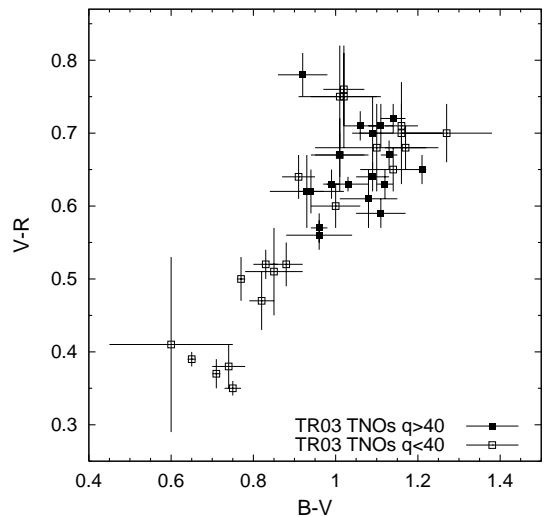}
      \caption[]{$B-V$ {\it vs.} $V-R$ plot of all TNOs from TR03
excluding the Centaurs. Without the Centaurs, the absence of
bimodality is clear.}
         \label{fig:teg03_all_tnos}
   \end{figure}
%- - - - - - - - - - - - - - - - - - - - - - - - - - - - - - - - - - -

%- - - - - - - - - - - - - - - - - - - - - - - - - - - - - - - - - - -
   \begin{figure}
   \centering
   \includegraphics[width=8.2cm, bb= 50 50 342 302]{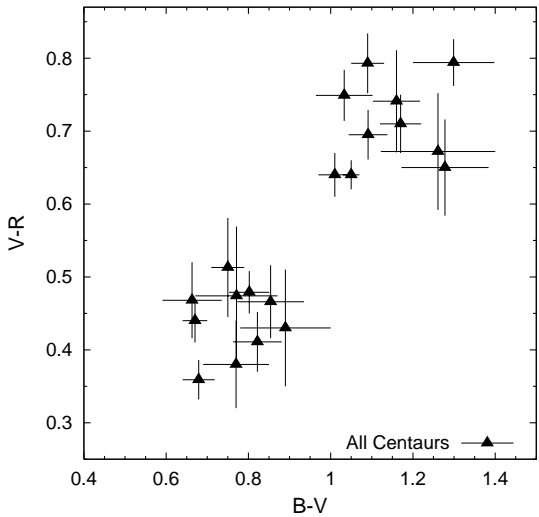}
      \caption{$B-V$ {\it vs.} $V-R$ plot of our analyzed Centaurs'
sample. The bimodality seems clear and the statistics come out highly 
significant.}
         \label{fig:all_cent_bv_vr}
   \end{figure}
%- - - - - - - - - - - - - - - - - - - - - - - - - - - - - - - - - - -

In the first work on TNOs bimodality (TR98), the two color groups
seemed very clear. However, statistics were performed on too few
objects to be really significant. Later, with an enlarged data sample,
the separation between the two presumed color groups was much more
tiny, even though still apparent (TR00). We should nonetheless notice
that in Tegler and Romanishin's analysis Centaurs were mixed with
TNOs. The irradiative and collisional environment of Centaurs should
be very different of that of TNOs, hence conclusions based on the
colors of both populations altogether should be taken with great care.
There are few statistical tests for the detection of more than one
mode (group) in a distribution. Tests based on bins, as used by TR03,
should be regarded with prudence as they are too dependent on the
bin's choice.
We will use only the Dip Test \citep{HarHar85,Har85}, a
distribution--free test that computes the maximum difference (``dip'')
between the empirical distribution function and the unimodal
distribution function that minimizes the differences.\\

If we first restrict ourselves to the 32 objects with known $B-R$ colors 
and $q<40~AU$ of TR03 --- a sample 
for which they found two groups ---, the dip test 
confirms a bimodality for $B-R$ colors with a significance
level ($SL$) of $99.7\%$. However, if we remove the Centaurs from this
sample we have bimodality only at $SL \sim 55\%$ for the 22 objects
left, while for the 10 Centaurs alone we see it at $SL=94.7\%$. Even
if the obtained $SL$ is too low and the number of objects is too small
to make strong conclusions, these results suggest that the bimodality
found is most probably dictated by the Centaurs colors and can not be
stated as general for TNOs (see figure \ref{fig:tr03_bimod}).
It should be noted also that, on TR03, these two groups were found
without the inclusion of the ``red cluster'' of TNOs with perihelia
distances above $40~AU$. These ``red cluster'' objects with $q>40~AU$
are also characterized by a low inclination orbit and they most
probably constitute a separate group \citep{LevStern01,Bro01} However,
they appear as a red cluster while compared to the full color
distribution of all other TNOs, that continuously range from neutral
to red. If TNOs with $q<40~AU$ divide in two groups one neutral and
another red, we can no longer consider the ``red cluster'' as a
separate group (based on color differences) since it is apparently
equal to the red one found for $q<40~AU$. Consequently, TR03 are
implicitly dealing, in their work, with not two but three groups of
objects: neutral and red TNOs with $q<40~AU$, and red TNOs with
$q>40~AU$.
When all objects are included in the analysis, 
the bimodality disappears ($SL\sim36\%$). 
If Centaurs are removed the
significance level goes down to $\sim9\%$ (see figures
\ref{fig:tr03_full_data} and \ref{fig:teg03_all_tnos}).  Thus, while it 
does not seem justifiable to ignore the ``red cluster'' in the
statistics, including those does ``destroy'' the evidence for
bimodality, even with the inclusion of Centaurs.
We will therefore investigate the possibility that the conclusion
for two color groups is in fact being misguided by the superposition
of a bimodal Centaur distribution over a continuous spread of colors
of TNOs with $q<40~AU$ and a separate group of exclusively red TNOs
with $q>40~AU$ or, globally, an unimodality for TNOs colors with an
excess of red objects.\\

We therefore took all Centaurs with available quasi--simultaneous
observations of $B-V$, $V-R$ and $B-R$ colors, building a data sample
of 20 objects which represents almost $50\%$ of all known
Centaurs. Eight objects were obtained under the ``ESO Large Program on
Centaurs and TNOs'', six from \cite{Pei03} and two from \cite{Boe02}
and another six drawn from the ``Meudon Multicolor Survey''
\citep{Dor02}. Six more were added from the compiled MBOSS database of
\cite{HaiDel02}, since their average color indexes were computed from
literature using only simultaneous magnitudes (see table
\ref{tab:all_centaurs}).
While TR03 observational methodology also monitored the absence of
significant magnitude variations, all their Centaurs were also
independently measured by the previously quoted works.
The dip test on our 20 objects data sample reveals bimodality for
$B-R$ colors with $SL=99.5\%$. Furthermore, by testing $B-V$ and $V-R$
colors separately, we see that the bimodal behavior is dominated by
the $V-R$ color index, since there is evidence for two $V-R$ color
groups at $SL=97.7\%$ while for $B-V$ alone it is inexistent. All dip
test results are summarized in table \ref{tab:dip_tests}.
We conclude that Centaurs' popu\-la\-tion is indeed composed by two
separate $B-R$ color groups with strong evidence for a dominating
$V-R$ bimodality.\\

In the color evolution model by \cite{Del03}, TNOs are resurfaced by a
competition of 1) reddening by irradiation; 2) bluishing by
non-disruptive collisions between TNOs; 3) bluishing by cometary
activity ({\em i.e.} uniform redeposition of neutral-colored dust
lifted by gas outbursts). Simulations using known Centaurs orbits lead
to very red objects regardless of the size. In the spaces between
giant planets, reddening by irradiation is the dominant process while
collisions are scarce; cometary activity outbursts (triggered by
collisions) are also possible but with a very low probability.
Observed blue Centaurs colors (like for 2060 Chiron) are compatible
with a surface being recovered by the bound coma detected by
\cite{Mee97}. It has been suggested that some short period comets and
Centaurs should be the fragments of objects ejected from the EKB
\citep{FD96}. Such fragmentation would expose some fresh, volatile
rich interior that would be compatible with the existence of active
and/or blue Centaurs. \cite{Del03} concluded that very red Centaurs
are old, fully irradiated objects (with a thick, dark irradiated
mantle that prevents from spontaneous -- {\em i.e.} non
collision-induced -- cometary activity outbursts), while blue centaurs
might be fresh fragments recovering from an ejection from the
EKB. Unfortunately, their model does not currently predict the
proportion of blue {\em vs} red Centaurs. However, following this
formalism, one can expect it is more likely to observe red Centaurs.
We see with the sample used that the two groups (red/blue) are equally
distributed. Assuming that the ``excess'' of blue objects is not an
observational bias, we will have to invoke an injection of new blue
Centaurs from the EKB.
\cite{YuTre99} suggested that evolution of Plutinos onto
Neptune--crossing orbit may dominate the flux of Jupiter--family
comets. Consequently, Centaurs' population may be mainly populated by
ex--Plutinos. An excess of blue Plutinos within the same size range as
Centaurs is reported by \cite{Pei03} as also similar color--color
correlations for Centaurs and Plutinos in opposition to SDOs
correlations. These results give plausibility to our hypothesis.
Nevertheless, different origins for the Centaurs population, or part
of it, as an explanation of the two color groups can not be discarded.

%-------------------------------------------------------------------------
   \begin{table}
     \begin{center}
      \caption[]{Centaurs' colors}
         \label{tab:all_centaurs}
        \setlength{\tabcolsep}{1.5mm}
         \begin{tabular}[]{llccc}
         \hline
         \hline
        & Object           & $B-V$         & $V-R$         & $B-R$ \\
         \hline
        & 2002 GO$_{9}$$^{\mathrm{a}}$    & $1.03 \pm 0.07$ & $0.75 \pm 0.04$ & $1.80 \pm 0.07$ \\ 
        & 2002 DH$_{5}$$^{\mathrm{a}}$    & $0.66 \pm 0.07$ & $0.47 \pm 0.05$ & $1.05 \pm 0.07$ \\ 
        & 1999 XX$_{143}$$^{\mathrm{a}}$  & $1.28 \pm 0.11$ & $0.65 \pm 0.07$ & $1.86 \pm 0.07$ \\ 
        & 1994 TA$^{\mathrm{b}}$          & $1.26 \pm 0.14$ & $0.67 \pm 0.08$ & $1.93 \pm 0.16$ \\
(63252) & 2001 BL$_{41}$$^{\mathrm{a}}$   & $0.82 \pm 0.06$ & $0.41 \pm 0.04$ & $1.18 \pm 0.06$ \\ 
(60558) & 2000 EC$_{98}$$^{\mathrm{c}}$   & $0.85 \pm 0.08$ & $0.47 \pm 0.05$ & $1.32 \pm 0.08$ \\ 
(55576) & 2002 GB$_{10}$$^{\mathrm{a}}$   & $1.09 \pm 0.05$ & $0.69 \pm 0.03$ & $1.77 \pm 0.05$ \\ 
(54598) & 2000 QC$_{243}$$^{\mathrm{d}}$  & $0.67 \pm 0.03$ & $0.44 \pm 0.03$ & $1.11 \pm 0.03$ \\ 
(52975) & Cyllarus$^{\mathrm{d}}$         & $1.17 \pm 0.05$ & $0.71 \pm 0.04$ & $1.88 \pm 0.05$ \\ 
(52872) & Okyrhoe$^{\mathrm{d}}$          & $0.89 \pm 0.11$ & $0.43 \pm 0.08$ & $1.32 \pm 0.11$ \\ 
(49036) & Pelion$^{\mathrm{c}}$           & $0.77 \pm 0.10$ & $0.47 \pm 0.10$ & $1.25 \pm 0.10$ \\ 
(44539) & 1999 OX$_{3}$$^{\mathrm{a}}$    & $1.16 \pm 0.06$ & $0.74 \pm 0.07$ & $1.89 \pm 0.06$ \\ 
(33128) & 1998 BU$_{48}$$^{\mathrm{d}}$   & $1.01 \pm 0.04$ & $0.64 \pm 0.03$ & $1.65 \pm 0.04$ \\ 
(31824) & Elatus$^{\mathrm{d}}$           & $1.05 \pm 0.02$ & $0.64 \pm 0.02$ & $1.69 \pm 0.02$ \\ 
(10370) & Hylonome $^{\mathrm{d}}$        & $0.77 \pm 0.08$ & $0.38 \pm 0.06$ & $1.15 \pm 0.08$ \\ 
(10199) & Chariklo$^{\mathrm{b}}$         & $0.80 \pm 0.05$ & $0.48 \pm 0.03$ & $1.28 \pm 0.06$ \\ 
(8405)  & Asbolus$^{\mathrm{b}}$          & $0.75 \pm 0.04$ & $0.51 \pm 0.07$ & $1.26 \pm 0.08$ \\ 
(7066)  & Nessus$^{\mathrm{b}}$           & $1.09 \pm 0.04$ & $0.79 \pm 0.04$ & $1.88 \pm 0.06$ \\ 
(5145)  & Pholus$^{\mathrm{b}}$           & $1.30 \pm 0.10$ & $0.79 \pm 0.03$ & $2.09 \pm 0.10$ \\ 
(2060)  & Chiron$^{\mathrm{b}}$           & $0.68 \pm 0.04$ & $0.36 \pm 0.03$ & $1.04 \pm 0.05$ \\ 
         \hline              
         \end{tabular}
%\begin{list}{}{}
%\item[$^{\mathrm{a}}$] Colors from \cite{Pei03}
%\item[$^{\mathrm{b}}$] Colors from \cite{HaiDel02}
%\item[$^{\mathrm{c}}$] Colors from \cite{Boe02}
%\item[$^{\mathrm{d}}$] Colors from \cite{Dor02}
%\end{list}
     \end{center}
{\footnotesize Colors are from, respectively, $\rm{^a}$ \cite{Pei03}, $\rm{^b}$
\cite{HaiDel02}, $\rm{^c}$ \cite{Boe02}, $\rm{^d}$ \cite{Dor02}} 
   \end{table}

%-------------------------------------------------------------------------
   \begin{table}
     \begin{center}
      \caption[]{Dip Test Results}
         \label{tab:dip_tests}
        \setlength{\tabcolsep}{1.5mm}
         \begin{tabular}[]{lcccr}
         \hline
         \hline
Data Sample  &n$^{\mathrm{a}}$& Color & Dip$^{\mathrm{b}}$& SL$^{\mathrm{c}}$ \\
         \hline
TR03                         &  &       &        &             \\
... Centaurs+TNOs:~$q<40~AU$   & 32 & $B-R$ & 0.1073 & $99.7\%$ \\
... Centaurs                   & 10 & $B-R$ & 0.1389 & $94.7\%$ \\
... TNOs:~$q<40~AU$        & 22 & $B-R$ & 0.0736 & $\sim 55\%$ \\
... Centaurs+all TNOs             & 50 & $B-R$ & 0.0442 & $\sim 36\%$ \\
... all TNOs                  & 40 & $B-R$ & 0.0411 & $\sim 9\%$  \\
         \hline
This work       & & & & \\
... Centaurs      & 20 & $B-R$ & 0.1211 & $99.5\%$ \\
... Centaurs      & 20 & $B-V$ & 0.0750 & $\sim 53\%$ \\
... Centaurs      & 20 & $V-R$ & 0.1156 & $97.7\%$ \\
         \hline              
         \end{tabular}
%%\begin{list}{}{}
%%\item[$^{\mathrm{a}}$] Number of objects
%%\item[$^{\mathrm{b}}$] Measure of minimum difference from unimodality
%%\item[$^{\mathrm{c}}$] Dip's significance level
%%\end{list}
     \end{center}
$\rm{^a}$ Number of objects, $\rm{^b}$ Measure of minimum difference from
unimodality, $\rm{^c}$ Dip's significance level
   \end{table}
%-------------------------------------------------------------------------

%%%%%%%%%%%%%%%%%%%%%

\section{Conclusions}

In this work, we revisited the TNOs color controversy with a different
and necessary twofold approach. First we argue that the dip test we
used is a more appropriate tool to investigate the uni- or bi-modal
distribution of colors. And second we suggest that the Centaur
population has not to be mixed with TNO population in statistical
studies since both populations are experiencing very different degrees
of surface processing. We combined new and published datasets to show
that the Centaur color distribution is bimodal while the TNO color
distribution is definitively unimodal.  As Centaurs are presumed
escapees from the EKB, this new scheme is still compatible with
evolution processes \citep{Del03}. Moreover, our results support a
Plutino origin for the Centaurs. However, given the uncertainties on
the transfer mechanisms of Centaurs from the EKB, we cannot rule out
the possibility of true compositional diversity of Centaurs as an
explanation of their color dichotomy.

\begin{acknowledgements}
N.P. ac\-know\-ledges funding from the Por\-tu\-guese Foundation for Science
and for Technology (FCT-SFRH:BD/1094/2000).
\end{acknowledgements}

\end{document}